\documentclass[twocolumn,showpacs,amsfonts,aps,prl,floatfix,nofootinbib,float,superscriptaddress]{revtex4-1} 
\usepackage{graphicx}  % needed for figures
\usepackage{dcolumn}   % needed for some tables
\usepackage{bm}        % for math
\usepackage{amssymb}   % for math
\usepackage{amsmath}   % for math
\usepackage{tabularx}

\newcommand{\beq}{\begin{equation}}
\newcommand{\eeq}{\end{equation}}
\newcommand{\bea}{\begin{align}}
\newcommand{\eea}{\end{align}}
\newcommand{\beas}{\begin{align*}}
\newcommand{\eeas}{\end{align*}}

\newcommand{\Tint}[1]{{\hbox{$\sum$}\!\!\!\!\!\!\!\int\,}_{\!\!\!\!\raise-0.9ex\hbox{$\scriptstyle{#1}$}}}

\begin{document}
\widetext

\title{Static quark-antiquark potential in the quark-gluon plasma from lattice QCD}
\author{Yannis Burnier}
\affiliation{
Institute of Theoretical Physics, EPFL, CH-1015 Lausanne, Switzerland}
\author{Olaf Kaczmarek}
\affiliation{Fakult\"at f\"ur Physik, Universit\"at Bielefeld, D-33615 Bielefeld, Germany}
\author{Alexander Rothkopf}
\affiliation{Institute of Theoretical Physics, Universit\"at Heidelberg, Philosophenweg 12,  D-69120 Germany  }

\date{\today}

\begin{abstract}
We present a state-of-the-art determination of the complex valued static quark-antiquark potential at phenomenologically relevant temperatures around the deconfinement phase transition. Its values are obtained from non-perturbative lattice QCD simulations using spectral functions extracted via a novel Bayesian inference prescription. We find that the real part, both in a gluonic medium as well as in realistic QCD with light $u$, $d$ and $s$ quarks, lies close to the color singlet free energies in Coulomb gauge and shows Debye screening above the (pseudo) critical temperature $T_c$. The imaginary part is estimated in the gluonic medium, where we find that it is of the same order of magnitude as in hard-thermal loop resummed perturbation theory in the deconfined phase.
\end{abstract}

\maketitle

The potential acting between a heavy quark and anti-quark in a thermal medium is a central ingredient in our understanding of the strong interactions, described by quantum chromo-dynamics (QCD). The bound states it sustains, heavy quarkonium, are precision probes connecting theory and experiment \cite{Brambilla:2010cs}. They allow us to test QCD via low temperature spectroscopy \cite{Gray:2005ur}, as well as through their in-medium modification \cite{Shuryak:1978ij,Asakawa:2003re,Jakovac:2006sf} observed in the quark gluon plasma created in relativistic heavy ion collisions. In particular the open question of melting and regeneration observed at RHIC and LHC \cite{Adare:2006ns} urges a quantitative understanding of their in-medium behavior. 

A wealth of intuition has been accumulated in the past based, in part, on analogies with Abelian theories \cite{Shuryak:1978ij}, potential modeling \cite{Nadkarni:1986as} and strong coupling approaches \cite{SC}. Lattice QCD at $T=0$ tells us \cite{Koma:2006si} that the potential rises linearly before flattening off due to string breaking. Perturbation theory on the other hand shows that Debye screening plays a major role in the deconfined phase. At $T\gtrsim T_c$, reached in current experiments, we expect that the medium gradually weakens the interaction. How the transition between the two regimes manifests itself quantitatively in the potential however remained unanswered. Due to recent conceptual and methods developments we are now able to present in this letter a first principles determination of the temperature dependence of the static inter-quark potential in the phenomenologically relevant, i.e. non-perturbative regime around the phase transition.

The advent \cite{Barchielli:1986zs} of modern effective field theory allowed to put the definition of the static potential on a rigorous mathematical footing. By exploiting the separation between the heavy quark rest mass and medium scales, a derivation from a dynamical QCD observable, the real-time thermal Wilson loop $W(t,r)$ was achieved,
\beq
V(r)=\lim_{t\to\infty} \frac{i\partial_t W(t,r)}{W(t,r)}\label{Eq:VRealTimeDef}.
\eeq
This expression has been evaluated at finite temperature in hard thermal loop (HTL) resummed perturbation theory \cite{Laine:2007qy} and was found to be complex valued. In the deconfined phase the real part shows Debye screening, while the imaginary part is related to the scattering (Landau damping) and absorption (singlet-octet transition) of gluons from the medium. Even though at leading order the real part coincides with the color singlet free energies in Coulomb gauge, this agreement is already not exact at next-to-leading order \cite{Burnier:2009bk}. 
Calculating the potential to higher order in perturbation theory is a difficult task \cite{Brambilla:2010vq} and given the size of the strong coupling and the infrared problems in gauge theories, %\cite{Linde:1980ts} 
it is evident that non-perturbative methods within QCD, such as lattice simulations are required. The main difficulty we face is that numerical calculations are performed in imaginary time without direct access to dynamical quantities, such as $W(t,r)$.

In Ref.~\cite{Rothkopf:2011db} a strategy was laid out how to evaluate the real-time definition Eq.~\eqref{Eq:VRealTimeDef} using Euclidean lattice QCD simulations. It is based on a spectral decomposition
\begin{eqnarray}
 \nonumber W(\tau)=\int d\omega e^{-\omega \tau} \rho(\omega)\,
\leftrightarrow\, \int d\omega e^{-i\omega t} \rho(\omega)= W(t),
\end{eqnarray}
where $W(\tau)$ denotes the Euclidean time Wilson loop accessible on the lattice.
The above can be combined with Eq.\eqref{Eq:VRealTimeDef} to yield
\begin{align}
\hspace{-0.2cm}V(r)=\lim_{t\to\infty}\int d\omega\, \omega e^{-i\omega t} \rho(\omega,r)/\int d\omega\, e^{-i\omega t} \rho(\omega,r), \label{Eq:PotSpec}
\end{align}
in turn relating the values of the potential to the spectral function $\rho(\omega,r)$, which can in principle be obtained from lattice QCD.

The first practical challenge lies in obtaining the function $\rho(\omega,r)$ in Eq.~(\ref{Eq:PotSpec}) from a finite lattice QCD dataset $W(\tau_n,r),~n=1..N_\tau$ with statistical errors. Extracting from it continuous spectral features is an inherently ill-posed problem, which however can be given meaning by the use of Bayesian inference. 
In this well established statistical approach, additional prior information is used to select a unique solution from an otherwise undetermined $\chi^2$ fit. Unfortunately the standard methods, such as the Maximum Entropy Method (MEM) or extended MEM \cite{Asakawa:2000tr} have been shown \cite{Burnier:2013fca} to be unreliable in reproducing the very narrow peak structures present in the spectrum. Only recently it has become possible to faithfully reconstruct the functional form of such spectral functions from Euclidean correlator data with the advent of an improved Bayesian approach detailed in Ref. \cite{Burnier:2013nla}. In particular it is devoid of the convergence problems inherent in the MEM.

The second challenge is related to the Fourier transform and the infinite time limit in Eq.~(\ref{Eq:PotSpec}). Ref.~\cite{Rothkopf:2011db} noted that the lowest lying spectral peak will dominate the late-time evolution and hence represents the potential contributions. A thorough understanding of the timescales influencing the shape of this spectral feature was however obtained only later in Ref.~\cite{Burnier:2012az}. It was established on general grounds that if a potential picture is applicable, we have to expect a skewed Lorentzian as the lowest lying peak in the spectrum
\begin{align}
&\rho\notag\propto\frac{|{\rm Im} V(r)|{\rm cos}[{\rm Re}{\sigma_\infty}(r)]-({\rm Re}V(r)-\omega){\rm sin}[{\rm Re} {\sigma_\infty}(r)]}{ {\rm Im} V(r)^2+ ({\rm Re} V(r)-\omega)^2}\\ \notag&+{c_0}(r)+{c_1}(r)({\rm Re} V(r)-\omega)+{c_2}(r)({\rm Re} V(r)-\omega)^2\ldots\,.
\end{align}
Its position and width correspond to the values of the real- and imaginary part respectively, which can be obtained from a fit to the reconstructed spectra (i.e. with $\sigma_\infty$ and $c_i(r)$ as parameters). The practicability of this strategy has been verified within HTL perturbation theory in \cite{Burnier:2013fca} and it was found that for the purpose of extracting the potential, the Wilson loop can possibly be replaced by other observables such as the Wilson line correlators in Coulomb gauge. One particular benefit of this observable is the absence of cusp divergences and a much improved signal to noise ratio on the lattice.

\begin{figure}[t!]
\centering
\includegraphics[scale=0.5, clip=true, trim=0 0.2cm 0 0]{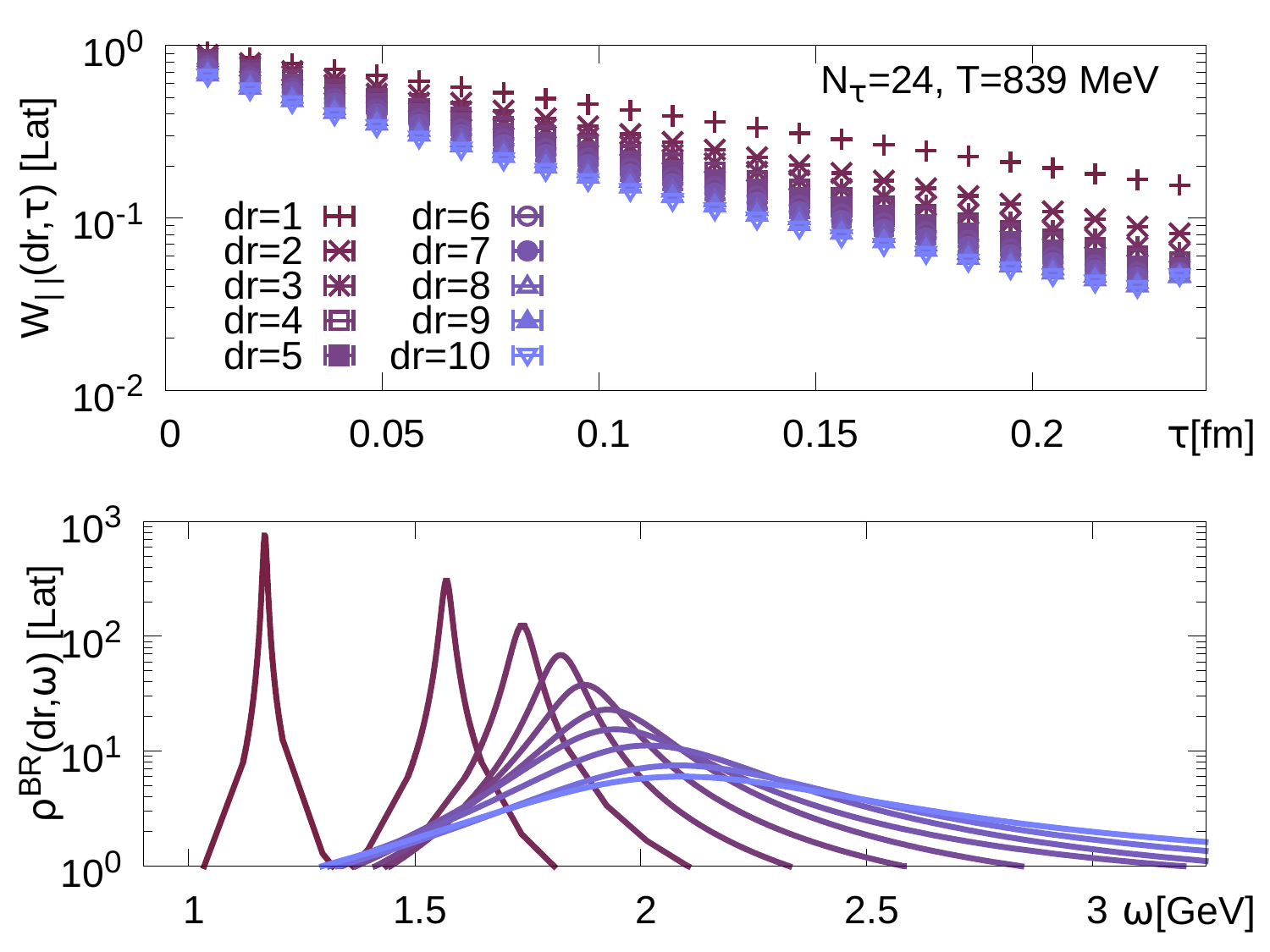}
\vspace{-0.3cm}
\caption{Spectral reconstruction: On-axis Wilson line correlator data (top) at $N_\tau=24$ and  (bottom) the spectral functions obtained by the new Bayesian reconstruction method.} 
\label{Fig1} \vspace{-0.4cm}
\end{figure}

Equipped with these technical and conceptual improvements we proceed to extract the temperature dependence of the static in-medium interquark potential in a purely gluonic as well as for the first time in a full QCD medium. We generated quenched QCD configurations, based on the naive anisotropic Wilson action in a fixed scale approach, i.e temperature is changed between $210{\rm MeV}(0.78T_c)\leq T\leq 839{\rm MeV}(3.11T_c)$ by modifying the number of temporal lattice points (see Tab.~\ref{Tab:LatParm}). Our choice of $\beta=7$ corresponds to a relatively fine lattice spacing of $a_s=0.039$fm \cite{Asakawa:2003re}, which together with a spatial extend of $N_s=32$ allows us to access both the Coulombic part of the potential, as well as those distances at which it is already screened at $3.11T_c$. We use an anisotropy of $a_s/a_\tau=4$, since for a reliable determination of ${\rm Im}V$, a large number of points in temporal direction is required. 

After fixing our configurations to Coulomb gauge, we measure the Wilson line correlators (see Tab.~\ref{Tab:LatParm} for the number of measurements $N_{\rm meas}$). Their values are obtained along each spatial axis (see e.g. top of Fig.~\ref{Fig1}), on the square- and cubic diagonals. Part of the finite lattice spacing artifacts are removed by correcting the spatial distances from a comparison of the free lattice propagator with the continuum \cite{Luscher:1995zz}.
 
\begin{table}[b!]
\vspace{-0.3cm}

\begin{tabularx}{8.8cm}{ | c | X | X | X | X | X | X | X | X | X | }
\hline
	SU(3):$N_\tau$ & 24 & 32 & 40 & 48 & 56 & 64 & 72 & 80 & 96 \\ \hline
	$T$[MeV] & 839 & 629 & 503 & 419 & 360 & 315 & 280 & 252 & 210 \\ \hline
	$N_{\rm meas}$ & 3270 & 2030 & 1940 & 1110 & 1410 &1520 & 860 & 1190 & 1800 \\ \hline
\end{tabularx}

\vspace{0.1cm}

\begin{tabularx}{8.8cm}{ | c | X | X | X | X | X | X | X| }
\hline
	QGP: $\beta$ \hspace{0.8cm}& 6.8 & 6.9 & 7 & 7.125 & 7.25 & 7.3 & 7.48 \\ \hline
	$T$[MeV] & 148 & 164 & 182 & 205 & 232 & 243 & 286 \\ \hline
	a [fm] & 0.111 & 0.1 & 0.09 & 0.08 & 0.071 & 0.068 & 0.057\\ \hline
	$N_{\rm meas}$ & 1295 & 1340 & 1015 & 840 & 1220 & 1150 & 1130 \\ \hline
\end{tabularx}
\caption{Lattice QCD configurations: (top) quenched SU(3) on $32^3\times N_\tau$ anisotropic $\xi_b=3.5$ lattices with $a_s=0.039$fm and  $T_c\approx271$MeV. (bottom) Isotropic HotQCD $48^3\times12$ lattices with asqtad action ($m_l=m_s/20,T_c\approx174$MeV).}\label{Tab:LatParm}
\vspace{-0.8cm}
\end{table}

We perform the Bayesian reconstruction of the Wilson line spectra at different temperatures excluding the first and last correlator data point at $\tau=0, \beta$ to avoid overlap divergences \cite{Berwein:2012mw}. In order to not introduce a bias for the functional form, we work with a flat default model $m(\omega)={\rm const}$. Frequencies are discretized at $N_\omega=4000$ along $\omega^{\rm num}\in[-168,185]\times N_\tau/24$ GeV with a $N_{\rm hr}=550$ high resolution subinterval around the lowest lying peak. This choice is large enough for the spectra to settle parallel to the default model at large $\omega$ and we have checked that further extending the $\omega$ range or the number of points does not change the outcome. A unique global solution is found based on an LBFGS minimizer with $512$bit precision arithmetic and a step size stopping criterion of $\Delta=10^{-60}$.
Several of the reconstructed spectra for $N_\tau=24$ are shown in Fig.~\ref{Fig1}.

\begin{figure}[t!]
\centering
\includegraphics[scale=0.5, clip=true, trim=0 0.2cm 0 0]{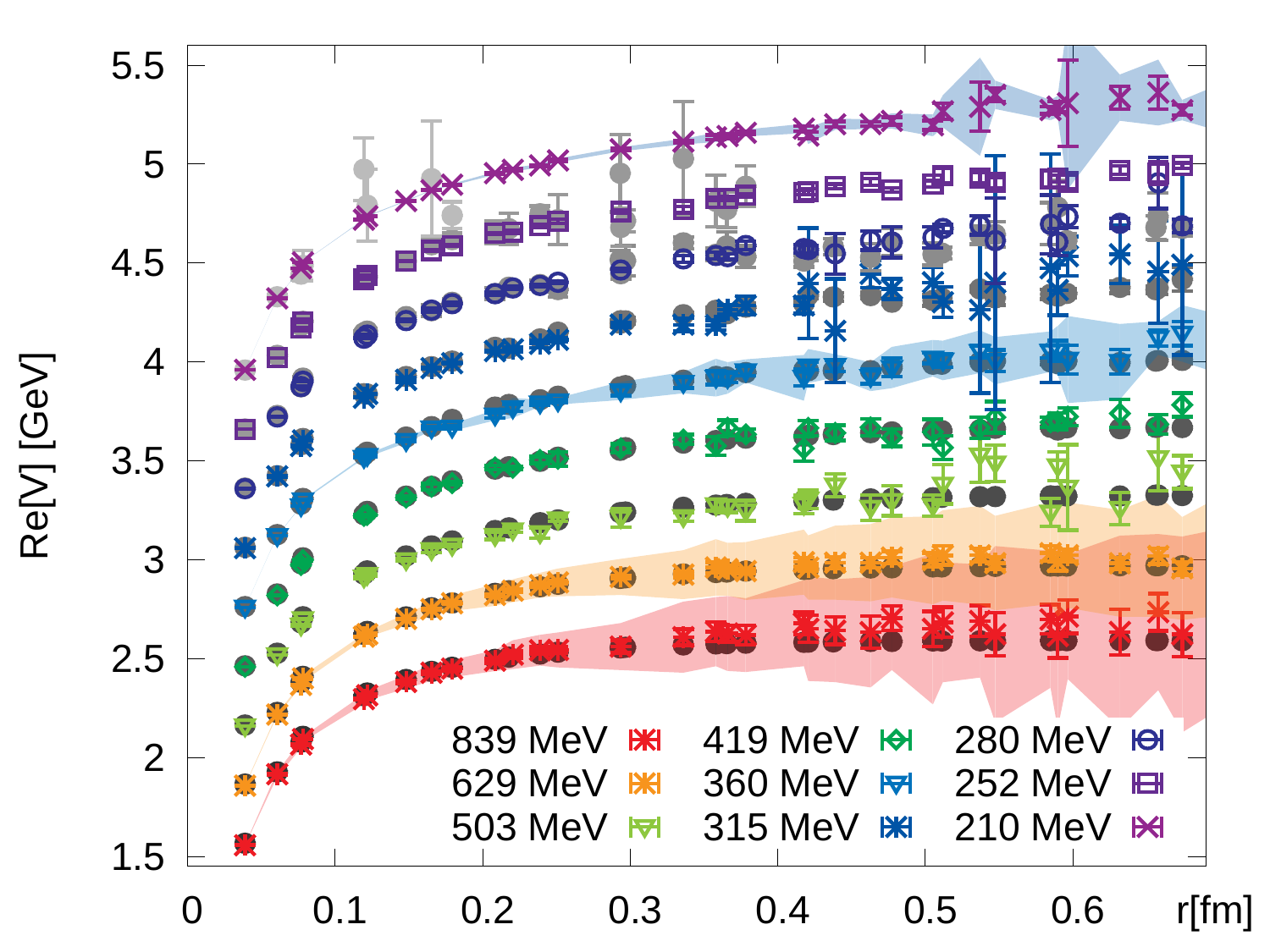}
\includegraphics[scale=0.5, clip=true, trim=0 0.2cm 0 0]{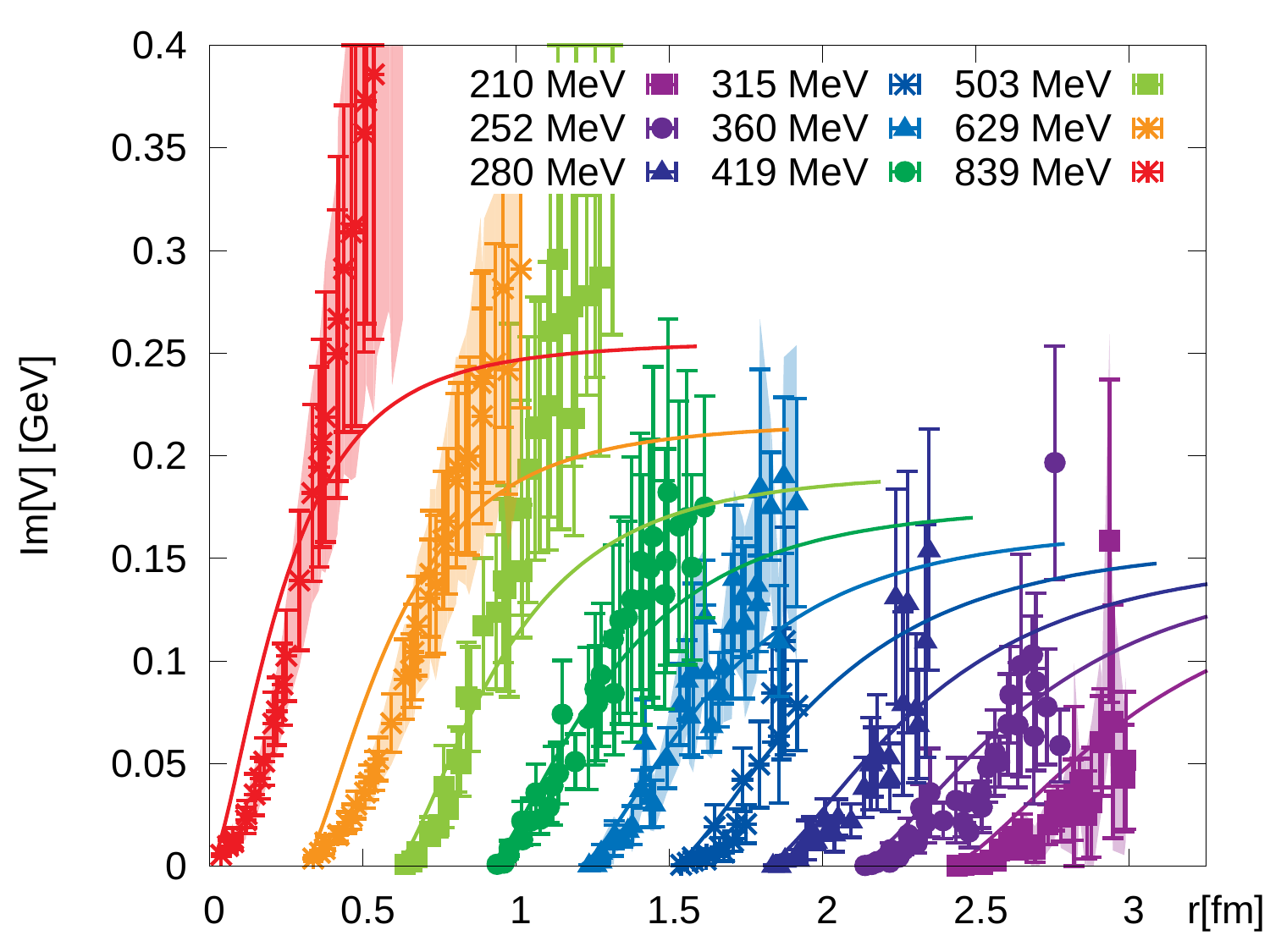}
\vspace{-0.2cm}
\caption{Gluonic medium: (top) The shifted real part of the static inter-quark potential (open symbols) compared to the color singlet free energies (gray circles). Error bars represent statistical uncertainty, error bands include also systematics (see main text). (bottom) ${\rm Im}[V]$ (symbols) shifted and compared to the HTL predictions (solid lines).} 
\label{Fig2} \vspace{-0.4cm}
\end{figure}

In the top panel of Fig.~\ref{Fig2} the results for the real part from the position of the lowest lying spectral peak are given by colored open symbols. They are contrasted to the color singlet free energies in Coulomb gauge $F^{(1)}(r)=-T{\rm log}[W_{||}(r,\tau=\beta)]$, obtained on the same lattices (filled gray circles). Since the raw values fall on top of each other at small distances we have shifted them for better readability. The error bars shown are obtained from the Jackknife variance resulting from repeating the reconstruction ten times excluding a different set of 10\% of the underlying measurements each. The error bands (given for $N_\tau=24,32,56,96$) on the other hand denote the maximum variance obtained from changing three different quantities. One corresponds to a reduction of the number of datapoints along $\tau$ by four and eight, the second to changing the default model normalization ($\times 10$, $\times 0.1$) or functional form ($m\propto{\rm const},\omega^{-2},\omega^2$) and the third 
to the reduction in signal to noise ratio by 
excluding 10\%,20\% or 30\% of the available measurements. Note that because the spectral reconstruction takes into account all datapoints along $\tau$, our results for $T \lesssim T_c$ are much more robust than the free energies, that rely on a single data point. On the other hand the Bayesian reconstruction suffers from a diminishing number of datapoints at increasing temperature, as seen in the errorbands. 

Our main observation is that even though the $\tau=\beta$ data point is excluded from the reconstruction, the values of ${\rm Re}[V]$ obtained at all temperatures lie close to the color singlet free energies. While the lowest temperature shows no or very weak deviation from a linearly rising potential, the values above $T>T_c$ show clear signs of Debye screening with increasing temperature. At $r<0.15$fm we find little temperature dependence, as expected. 

The extraction of the imaginary part from Bayesian spectra poses an even more formidable challenge than ${\rm Re}[V]$. Its presence can be qualitatively inferred already from the Euclidean correlator (see Fig.~\ref{Fig1} top panel), where at intermediate $\tau$ values a deviation from the exponential decay and a finite curvature emerges.
To obtain quantitative results, the reconstruction of the lowest lying peak needs to capture not only the width, which encodes ${\rm Im}[V]$ but also the overall skewed Lorentzian shape related to non-potential effects. 

The novel Bayesian approach for the first time allows us to extract this functional form (see Fig.~\ref{Fig1} bottom panel), where the MEM yielded Gaussian like features. Previous tests based on mock data from momentum regularized HTL show that to obtain values accurate to $\sim 25$\%, datasets with $N_\tau\sim{\cal O}(100)$ datapoints are required at a high precision of $\Delta D/D<10^{-4}$. 
If less points are available the reconstruction tends to underestimate the width, while statistical noise leads to broadening. The former effect dominates at high temperatures and at small separation distances $r<0.25$fm where the lattice data carries small relative errors, while at larger distances the exponential suppression of the Euclidean correlator leads to an artificially broad width. 

Taking these systematic effects into account, we can estimate the values of ${\rm Im}[V]$ to a lie in a band which is compatible with the expectations from HTL perturbation theory at high temperature and appears to lie slightly below HTL at low temperature.
The results of our first-principles investigation are consistent with the findings by a recent modeling approach based on HTL spectral functions \cite{Bazavov:2014kva}.

\begin{figure}[t!]
\centering
\includegraphics[scale=0.5, clip=true, trim=0 0.2cm 0 0]{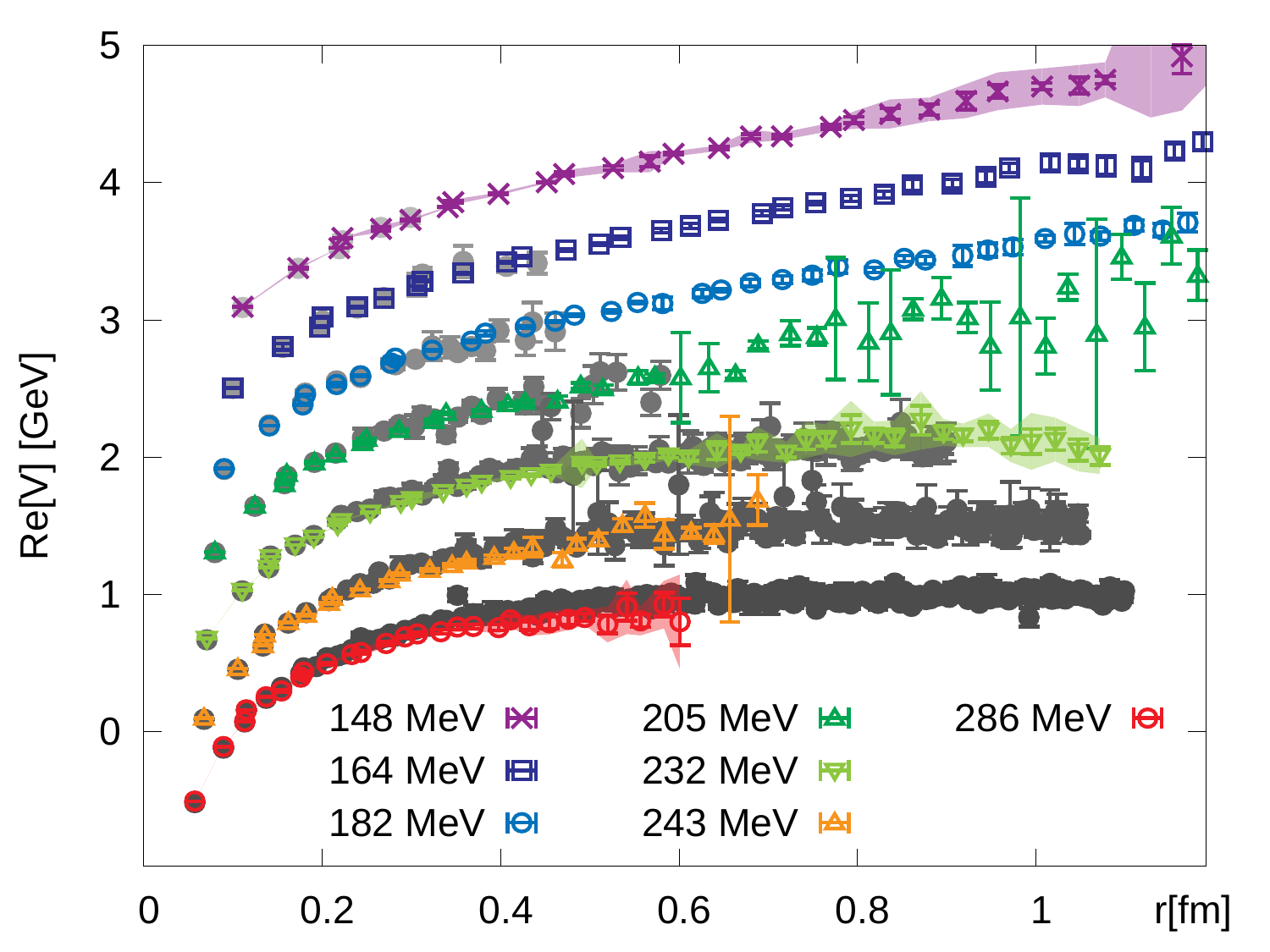}
\vspace{-0.3cm}
\caption{Quark-gluon-plasma: The real part of the static interquark potential (open symbols) compared to the color singlet free energies in Coulomb gauge (gray circles).} 
\label{Fig03} \vspace{-0.5cm}
\end{figure}

Next we consider the realistic setting of a thermal QCD medium containing both gluons, as well as the light $u,d$ and $s$ quarks. The corresponding full QCD $48^3\times12$ lattices were generated by the HotQCD collaboration \cite{Bazavov:2011nk} for the study of the QCD phase structure (see Tab.\ref{Tab:LatParm}). The Bayesian reconstructions with a common $\beta^{\rm num}=20$ are performed using $N_\omega=4600$ steps in a numerical interval of fixed length $\omega\in[-11,12]$ and a high resolution interval of $N_{\rm hr}=1000$ points to capture the lowest lying peak. 
Due to the high cost in generating the configurations it is currently not possible to obtain similarly large temporal extends as in quenched QCD, even with the use of supercomputers. Therefore we focus  in Fig.~\ref{Fig03} solely on the values of the real-part (colored symbols) of the potential, which are compared to the color singlet free energies (gray circles) from the same lattices. Error bars are again obtained from Jackknife variance. The error bands ($\beta=6.8,7.25,7.48$) result from the maximum variation among changing the number of datapoints along $\tau$ by one and two, changing the normalization and functional form of the default model as well as from removing 10\%,20\% or 30\% of the statistics.

At temperatures below and slightly above the pseudo-critical temperature $T_c\approx174$MeV on our lattices, the Bayesian reconstruction allows us to reliably determine ${\rm Re}[V]$ up to physical distances of $r=1$fm. The signal of the free energies at similar T is quickly lost in the much larger statistical noise. We do not observe string breaking, up to $r<1.2$fm, most probably due to the lattice pion mass $M^{\rm RMS}_\pi\sim300$MeV still lying above the physical value. The presence of fermionic d.o.f. significantly changes the character and location of the phase transition. Debye screening is already pronounced at the $T=286$MeV. Fig.~\ref{Fig03} shows that just as in the quenched case the real part lies close to the singlet free energy.  

We have measured the static in-medium inter-quark potential defined from first principles in lattice QCD simulations at phenomenologically relevant temperatures around the deconfinement transition. The real part of the complex potential is found to lie close to the color singlet free energies in Coulomb gauge and displays the expected Debye screened behavior above the phase transition. Consequently it disagrees with the internal energies, another observable that was used as potential in phenomenological studies. It would be hence interesting to use this first principles potential as input to heavy quarkonium spectral function and phenomenological studies \cite{Burnier:2007qm}. Our estimate of the imaginary part of the potential in the quenched case also shows reasonable agreement with HTL above $T_c$. At $T<T_c$ it is systematically smaller, as can be expected from the Boltzmann suppressed density of pions and glue balls present there.

While the spacing of our quenched lattices is already fine, a true continuum extrapolation needs to be pursued in the future. Some conceptual hurdles exist. It is long known that the Wilson loop does not possess a continuum limit due to cusp divergences \cite{Korchemsky:1987wg}. The Wilson line correlators contain end-point divergences, which accidentally vanish in Coulomb gauge up to two loops \cite{Aoyama:1981ev} but their non-perturbative behavior towards $a\to0$ is unknown. 

A more accurate reconstruction of ${\rm Im}[V]$ will also require lattices of significantly higher temporal resolution, which is currently challenging for quenched and impractical for full QCD. A concerted effort towards the tuning of dynamical QCD lattices with anisotropy would thus be greatly beneficial to this field of research.

The authors thank H.~B. Meyer, M.~P. Lombardo, P. Petreczky and J.-I. Skullerud for fruitful discussions. YB is supported by SNF grant PZ00P2-142524 and AR acknowledges partial support by the DFG.

\end{document}